\documentclass[fleqn,10pt,twocolumn]{wlscirep}
\usepackage[utf8]{inputenc}
\usepackage[T1]{fontenc}
\title{Compact ultrafast intense LWFA-driven crystal-based source of $\gamma$-radiation, positrons and neutrons}

\author[1,*,+]{Alexei Sytov}
\author[2,+]{Hyeon Myung Jeong}
\author[3]{Leejin Bae}
\author[4]{Iryna Chaikovska}
\author[2]{Hyoeun Choi}
\author[1]{Gianfranco Patern\`o}
\author[1,*,$\dagger$]{Laura Bandiera}
\author[2,*,$\dagger$]{Inhyuk Nam}

\affil[1]{INFN Ferrara Division, Ferrara 44122, Italy}
\affil[2]{Department of Physics, Ulsan National Institute of Science and Technology, Ulsan 44919, Republic of Korea}
\affil[3]{Gwangju Institute of Science and Technology, Institute for Basic Science, Gwangju 61005, Republic of Korea}
\affil[4]{Universit\'e Paris-Saclay, CNRS/IN2P3, IJCLab, Orsay 91400, France}

\affil[*]{sytov@fe.infn.it, bandiera@fe.infn.it, ihnam@unist.ac.kr}

\affil[+]{these authors contributed equally to this work}
\affil[$\dagger$]{these authors contributed equally as senior authors}

\begin{abstract}

Compact sources of intense $\gamma$-rays, positrons and neutrons can bring radiation and particle-beam capabilities to small university-scale laboratories that would otherwise rely on large accelerator facilities. Here we propose and simulate a compact, laser-driven source that combines laser plasma wakefield acceleration (LWFA) with oriented crystalline targets to enhance the conversion of relativistic electron beams into high-energy photons and secondary particles. Electrons aligned with major crystallographic directions undergo coherent interactions, including channeling radiation and coherent bremsstrahlung, producing substantially enhanced photon emission compared with amorphous targets. Moreover, the strong angular-spectral correlation of the emitted radiation enables the generation of collimated $\gamma$-ray beams with reduced spectral bandwidth. Consequently, crystal-enhanced emission can increase particle-production efficiency, namely that of positrons through $\gamma$ conversion into electron-positron pairs and of neutrons through photonuclear reactions. Using Geant4 simulations, we investigate electron energies of 300 MeV, 1 GeV and 3 GeV interacting with an oriented tungsten crystal. Crystal orientation increases the $\gamma$-ray, positron and neutron yields by up to a factor of $\sim2$, reaching production rates of $\sim10^{24}\gamma/s$, $\sim10^{23}e^+/s$ and $\sim10^{21}neutrons/s$ for a 200 pC electron bunch with a duration of a few fs. The collimated $\gamma$-ray brightness reaches $\sim10^{24}\gamma/s/mm^2/mrad^2/0.1\%BW$, with an enhancement of 6-8 compared with random crystal alignment. We further demonstrate tunable quasi-monochromatic radiation through coherent bremsstrahlung in a thin diamond crystal, reaching a brilliance of $\sim3.5\cdot10^{20}\gamma/s/mm^2/mrad^2/0.1\%BW$. Furthermore, we discuss the potential applications of the technique proposed. 

\end{abstract}
\begin{document}

\flushbottom
\maketitle
%
%
\thispagestyle{empty}


\section*{Introduction}


Laser wakefield acceleration (LWFA) has emerged as a promising route towards compact electron accelerators\cite{Tajima1979,Esarey2009}, with the potential to reduce both facility size and cost and thereby broaden access to relativistic electron beams. Whereas conventional accelerators provide $\gamma$-ray, positron and neutron beams with useful energies and intensities at only a few large-scale facilities with limited beam time, LWFA can generate such beams at high-power laser facilities. Since the first demonstrations of quasi-monoenergetic electron beams at energies of order 100 MeV\cite{Mangles2004,Geddes2004,Faure2004}, the maximum electron energy has increased to 1 GeV\cite{Leemans2006}, 8 GeV\cite{Gonsalves2019} and approximately 10 GeV\cite{Aniculaesei2024,Picksley2024}.
In parallel, controlled injection schemes, including colliding-pulse\cite{Faure2006}, ionization-induced\cite{McGuffey2010,Pak2010} and density down-ramp or shock-front injection\cite{Geddes2008,Schmid2010,Buck2013}, have enabled control of the beam energy, charge and energy spread, and beam loading has raised the bunch charge to the nanocoulomb level\cite{Couperus2017}. As a result, LWFA can now deliver relativistic electron beams with tunable parameters from compact laser systems.   
The most demanding applications of LWFA, such as free-electron lasers operated at 27 nm\cite{Wang2021} or in seeded mode\cite{Labat2023,Labat2026} and injectors for synchrotron light sources\cite{Antipov2021,Winkler2025}, impose stringent requirements on beam emittance, energy spread, stability and reproducibility. By contrast, target-based generation of secondary radiation and particles can often tolerate less stringent beam-quality requirements, because the total yield is governed primarily by the electron energy, bunch charge, angular distribution and repetition rate.

LWFA beams are therefore attractive drivers for compact target-based sources of high-energy radiation and secondary particles\cite{Andronic2026}. Whereas betatron X-rays are emitted directly in the plasma wake\cite{Rousse2004,Kneip2010,Corde2013,Albert2023} and inverse Compton $\gamma$-rays require a second laser pulse\cite{Phuoc2012,Sarri2014,Tsai2026}, a high-Z converter placed downstream of the accelerator converts LWFA electrons into short-pulse bremsstrahlung $\gamma$-rays, positrons, neutrons and muons\cite{Glinec2005,Lemos2018,Sarri2013,Song2023,Noh2024,Vallieres2025,Kim2026,Zhang2025Muon,Terzani2025}. Such $\gamma$-ray beams are relevant to photonuclear studies, nuclear resonance fluorescence, high-energy radiography, non-destructive inspection and radiation-damage testing of aerospace materials and electronics\cite{Glinec2005,Lemos2018}, and bremsstrahlung-driven photonuclear reactions populate nuclear isomers for reaction-data and medical-isotope studies\cite{Lan2023,Feng2023,Giubega2025}. Positron beams are used for materials characterization by positron annihilation spectroscopy and are considered as injectors for future colliders\cite{Audet2021,Alejo2024,Alharthi2025}, and neutron beams support nuclear physics, neutron imaging, activation analysis and isotope production\cite{Feng2020,Vallieres2025,Kim2026}. Generating these complementary probes from an LWFA electron beam interacting with a target is therefore attractive for both scientific and industrial applications.

\begin{figure*}[ht]
\centering
\includegraphics[width=0.83\textwidth]{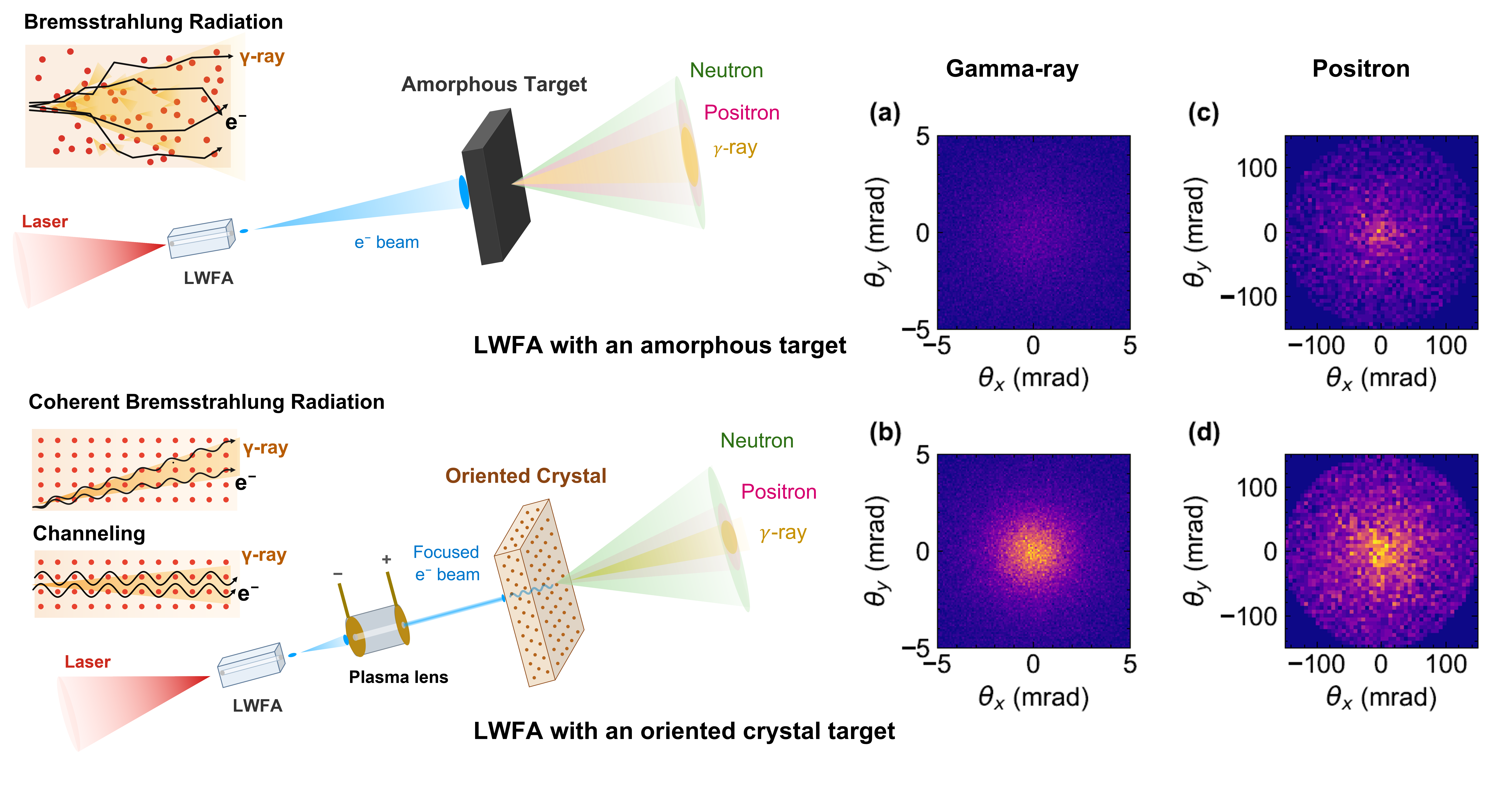}
\caption{Schematic of the proposed setup coupling an LWFA electron source to an amorphous target (top) and a plasma lens and an oriented crystal (bottom). Geant4 simulation results: 2D distribution of (a,b) gamma-ray and (c,d) positron for LWFA with an amorphous target and with a plasma lens and oriented crystal target, respectively. }
\label{Fig1}
\end{figure*}

However, the conversion efficiency of conventional amorphous targets limits the secondary-particle and photon yields achievable with a given electron beam. Increasing the yield generally requires higher beam power or thicker targets, which increases energy deposition and can lead to severe thermal loads and target damage. A thicker target also enhances multiple Coulomb scattering, which broadens the angular distribution and enlarges the effective source size of the emitted particles, thereby degrading the emittance and brightness of the secondary beams\cite{Alejo2024}. Moreover, the emitted radiation typically has a broad energy spectrum, limiting applications that require high spectral selectivity.

Oriented crystals have been extensively studied with conventional accelerator beams and have demonstrated their potential as intense sources of secondary particles and high-energy photons\cite{Baier1998,Uggerhoj2005,Bandiera2025,Bandiera2023,Bandiera2022Pair}. Coherent effects arise from the collective electromagnetic interaction of a charged particle with crystallographic planes or atomic strings when it propagates at a sufficiently small angle with respect to them. The particle then experiences a strong, spatially periodic electric field of an ordered crystal lattice.  This interaction can induce oscillatory motion and substantially increase the photon yield compared with amorphous materials or randomly oriented crystals, in analogy with radiation generation in magnetic undulators and wigglers. Depending on the incidence angle with respect to the crystal axes or planes, this enhancement arises mainly from two mechanisms, channeling radiation and coherent bremsstrahlung.


\textit{Channeling} occurs when charged particles enter a crystal at a sufficiently small angle with respect to crystallographic planes or axes and become confined by the corresponding transverse electric field\cite{Kumakhov1976,Baier1998,Sytov2019}. The particles consequently undergo transverse oscillations while propagating through the crystal and emit radiation, referred to as \textit{channeling radiation}. Depending on the particle energy, crystal material, crystallographic orientation and channeling conditions, the radiation can exhibit different spectral regimes, spanning from dipole-like emission with distinct harmonics, favorable for quasi-monochromatic photon beams, to more intense, synchrotron-like broadband emission in stronger-field regimes. Experimentally, coherent radiation associated with channeling and strong-field interactions in oriented crystals has been studied over a broad energy range, from electron energies of a few MeV up to several hundred GeV, corresponding to photon energies from a few keV to hundreds of GeV\cite{Bandiera2021,Bandiera2023,Bandiera2025}.

\textit{Coherent bremsstrahlung} (CB) occurs when a charged particle traverses an oriented crystal without being confined in channeling, while still experiencing the periodic electric field of the crystal planes\cite{Baier1998}. This periodicity leads to constructive interference of the emitted radiation at selected photon energies, producing pronounced spectral peaks superimposed on the ordinary bremsstrahlung spectrum. The position of these peaks can be tuned by changing the crystal alignment with respect to the incident beam, providing control over the energy of the enhanced photon emission. CB has been exploited at facilities such as MAMI in Mainz and Jefferson Lab to produce intense, energy-selectable and linearly polarized photon beams for photonuclear and photoproduction experiments\cite{Lohmann1994,Bandiera2021}.

The enhanced radiation produced in an oriented crystal can be exploited for secondary-particle generation, either within the crystal itself or in a downstream converter target. The underlying conversion processes are the same as in amorphous materials, but benefit from the enhanced photon yield provided by coherent crystal effects. Secondary electrons and positrons produced at small angles to crystallographic planes or axes can also undergo coherent interactions, which enhance electromagnetic shower development and can modify particle angular distributions.
Electron–positron pair production by photons generated in oriented crystals has been extensively studied as a route towards intense positron sources\cite{Bandiera2022Pair,Soldani2024,Bandiera2025} and has been proposed for the FCC-ee injector\cite{Alharthi2025}.

In this work, we propose a compact source of intense radiation and secondary particles based on the combination of LWFA and an oriented crystal. We consider electron beams with energies of 300 MeV, 1 GeV, and 3 GeV, representative of those achievable with modern tens-of-TW to PW-class laser systems. Because the few-mrad divergence typical of LWFA beams exceeds the angular acceptance of the crystal, a plasma lens placed after the accelerator collimates the beam before it reaches the crystal. The beams are then incident on a W crystal aligned along the $\langle111\rangle$ axis, where coherent interactions enhance the production of high-energy photons. Using the Geant4 toolkit and the G4ChannelingFastSimModel\cite{Agostinelli2003,Sytov2023,Geant4Manual}, we simulate the resulting $\gamma$-ray, positron and neutron yields and compare them with those obtained for a randomly oriented W target. The corresponding energy spectra and angular distributions are analyzed to characterize the source performance. We also simulate $\gamma$-radiation produced in a diamond crystal through the coherent bremsstrahlung regime at different orientations with respect to a crystal plane and estimate the brilliance of radiation within quasi-monochromatic radiation peaks. We also discuss the possibilities of using different coherent radiation regimes in oriented crystals at other beam energies.

\section*{Results}
\subsection*{Setup description}
The proposed setup coupling an LWFA electron source to an oriented crystal is schematically shown in Fig. \ref{Fig1}. In the typcial configuration [Fig. \ref{Fig1}(top)], an LWFA electron beam is transported directly to a target to generate secondary beams. Despite its intrinsically micron-scale source size, the beam typically has a divergence of a few milliradians, causing it to expand to several hundred micrometers after propagating a few tens of centimeters. Here, we propose a scheme in which a plasma lens is placed immediately after the accelerator stage [Fig. \ref{Fig1}(bottom)].

A plasma lens can be used to focus the electron beam and reduce its angular divergence to meet the requirements for coherent interactions in the crystal\cite{Lehe2014,Thaury2015,Schmid2016,Tilborg2015,Integrated2023}. The oriented crystal then acts as a radiator, converting the incident electron beam into an intense photon beam. The generated photons can subsequently produce secondary particles either within the same crystal or in a separate downstream converter target.

The lens refocuses the beam before it reaches the target, allowing the small LWFA source size to be preserved while satisfying the low-divergence requirement for efficient interaction with an oriented crystal. Under these conditions, the oriented crystal can enhance the secondary-beam yield while reducing both beam divergence and target length. The shorter target suppresses Coulomb scattering, thereby improving the quality of secondary beams such as gamma rays, positrons, and neutrons.

Consequently, the secondary beams can exhibit substantially higher peak brightness and improved transverse coherence while retaining an extremely small source size. Our simulations demonstrate that these intrinsic advantages of LWFA beams can be leveraged to markedly enhance secondary-beam performance.

To investigate the proposed concept, we use the simplified geometry illustrated in Fig. \ref{Fig1}, implemented in Geant4\cite{Agostinelli2003}. Particle interactions with the oriented crystal are simulated with the G4ChannelingFastSimModel\cite{Sytov2023,Geant4Manual}, which is described in detail in \textbf{Methods}. In Geant4, the electron beam is initialized at the crystal surface with parameters representative of an LWFA beam after transport and focusing by a plasma lens. The beam parameters used in the simulations are summarized in Table \ref{Table0} for 3 different beam energies, while their justification is discussed below. All the particles escaping the target through its back surface were recorded including their relevant kinematic properties. For comparison, the coherent crystal effects can be disabled in the simulation, to treat the same target as an amorphous material. The Geant4 physics includes the relevant electromagnetic and photonuclear processes responsible for $\gamma$-ray, positron and neutron production and transport.

\begin{table*}[ht]
\centering
\begin{tabular}{|l|l|l|l|l|l|l|l|}
\hline
electron & energy & r.m.s. beam size & r.m.s. angular diver- & FWHM bunch & $\theta_L$ (mrad)& $\theta_v$ (mrad) & $\theta_c$ (mrad)\\
beam energy & spread & x (y) ($\mu$m) & gence x (y) (mrad) & pulse duration & W $\langle111\rangle$ & W $\langle111\rangle$ &   \\
\hline
300 MeV & 1 \%  & 3 (3) & 0.4 (0.4) & 10 fs &2.4 & 1.7 & 1.7 \\
\hline
1 GeV & 1 \%   & 3 (3)  & 0.4 (0.4) & 10 fs &1.3 & 1.7 & 0.51\\
\hline
3 GeV & 1 \% & 2 (2) & 0.4 (0.4) & 10 fs &0.77 & 1.7 & 0.17 \\
\hline
\end{tabular}
\caption{\label{Table0}Electron beam parameters at the crystal entrance and characteristic angles of coherent effects in a crystal.}
\end{table*}

\begin{figure}[ht]
\includegraphics[width=\columnwidth]{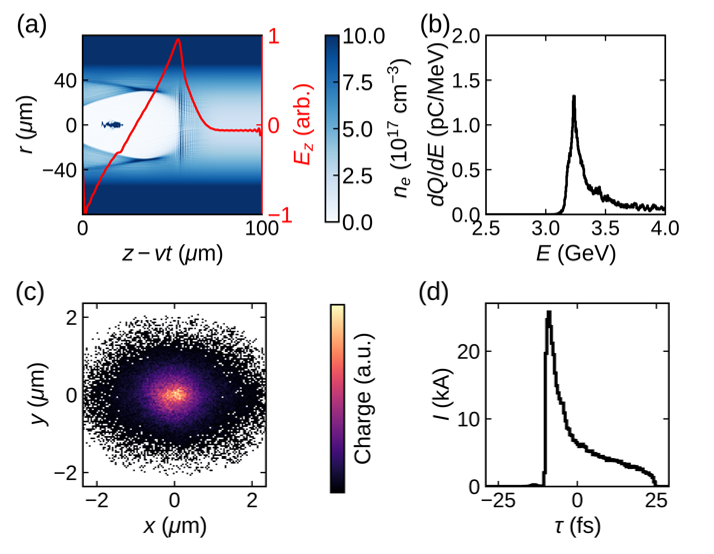}
\caption{Simulation results of electron beam from a laser wakefield acceleration. (a) Electron density after the laser pulse has propagated 100 mm. The red line is the wakefield $E_z$. (b) Energy spectrum. The central energy is $E_c=3.25$ GeV, the energy spread is 1.85 \% (r.m.s.) and the bunch charge is 246 pC. (c) Transverse beam charge distribution with a r.m.s. beam sizes of 0.76 $\mu$m in $x$ and 0.53 $\mu$m in $y$ , respectively. (d) Beam current profile with a bunch pulse duration of 4.6 fs (FWHM).}
\label{FigLWFA}
\end{figure}

\begin{figure}[ht]
\centering
\includegraphics[width=\columnwidth]{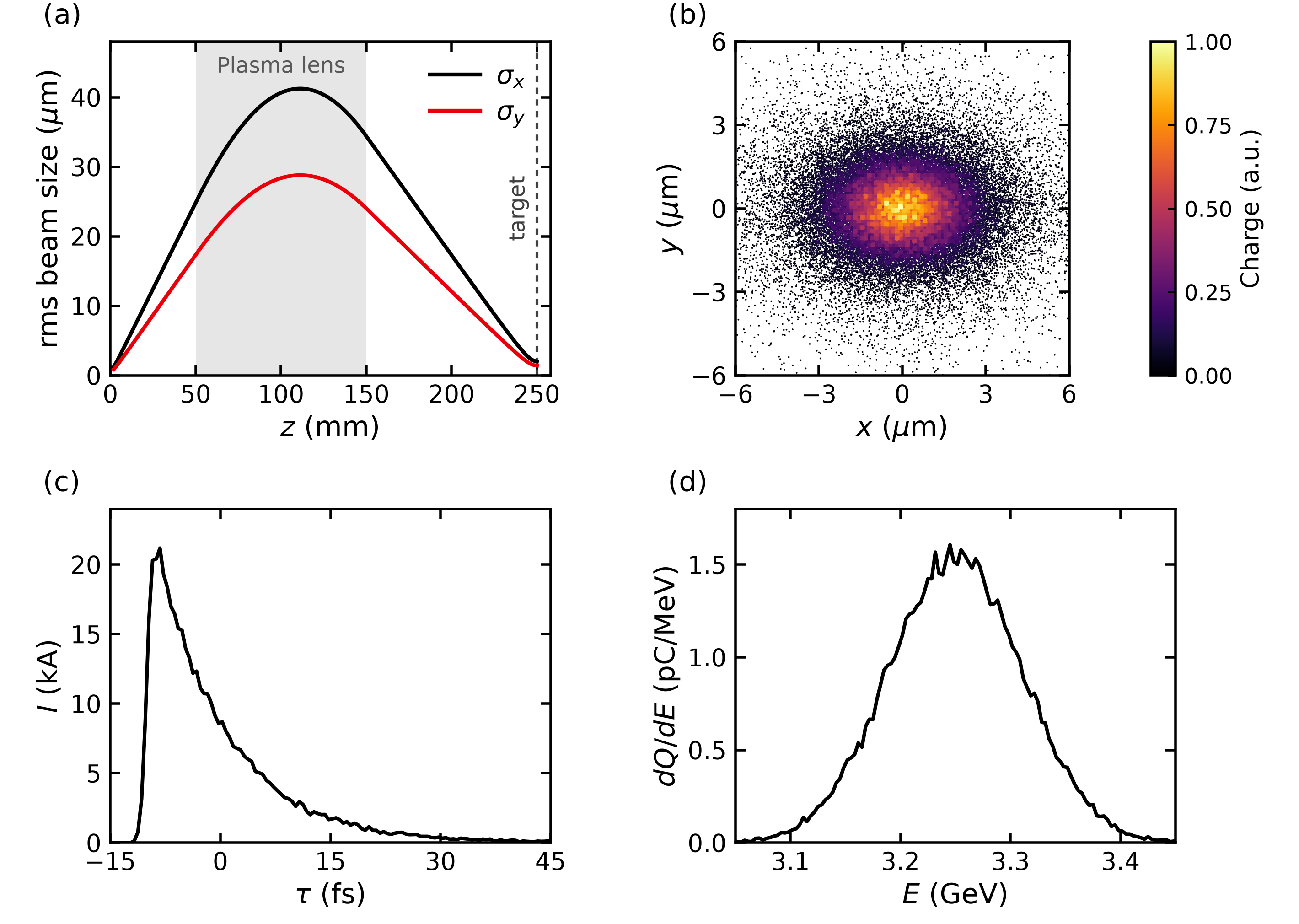}
\caption{Transport of the LWFA electron beam from the plasma exit to the crystal through an active plasma lens and beam properties at the entrance of crystal. (a) Evolution of the r.m.s. beam sizes $\sigma_x$ and $\sigma_y$ along the beamline. The shaded area marks the 100 mm long plasma lens, which starts 50 mm downstream of the plasma exit and operates at a discharge current of 771 A, corresponding to a focusing gradient of 2.5 kT m$^{-1}$. The dashed line marks the crystal position 250 mm from the source. The r.m.s. beam size at the crystal is 2.02 $\mu$m and 1.41 $\mu$m and the r.m.s. divergence is 344 $\mu$rad and 240 $\mu$rad in $x$ and $y$, respectively (b) Transverse charge density. (c) Current profile at the crystal. The bunch pulse duration increased to ~8.6 fs (FWHM) (d) Energy spectrum with an r.m.s energy spread of 1.85 \%. }
\label{FigAPL}
\end{figure}

\subsection*{LWFA electron beam parameters}

The three beam energies considered here are routinely reached in laser wakefield experiments at intense laser facilities. Petawatt pulses at CoReLS produce stable, high-quality beams at energies reaching 3 GeV\cite{Mirzaie2024}, similar energy of 2 GeV is routinely provided at ELBA ELI Beamlines \cite{ELI} with possible extension to 5 GeV, and the BELLA laser reached 8 GeV in a laser-heated capillary discharge waveguide\cite{Gonsalves2019} and 9.2 GeV, with charge extending beyond 10 GeV, in a 30 cm hydrodynamic optical-field-ionized channel\cite{Picksley2024}.

The bunch charge is governed by the injection scheme. Controlled injection such as density down ramp and ionization injection delivers tens of picocoulombs with percent-level energy spread\cite{McGuffey2010,Zeng2014}, whereas beam-loaded operation raises the charge to the nanocoulomb scale at the cost of a broader spectrum\cite{Couperus2017}. Down ramp injection tunes the charge, energy and energy spread continuously between these limits\cite{Schmid2010,Buck2013,Hue2023,Cobo2024,Dopp2018}. 
To verify that a single beam meets the energy, charge and divergence requirements of the crystal target at the same time, we simulated the chain from injection to the crystal entrance with a particle-in-cell code coupled to a beam transport code. The numerical setup is given in \textbf{Methods}.

A 200 TW-class driver with $a_0=2.0$, $w_0=45~\mu$m, $\tau=40$ fs and $\lambda_0=800$ nm is focused at $z_f=4.5$ mm into a 100 mm long preformed plasma channel. The channel is matched to a 50 $\mu$m spot size in the first density plateau and to 30 $\mu$m in the second one, and the simulated spot size stays within a converged range around 27 $\mu$m over the full channel length. Electrons are injected at a density down ramp located between 4.5 mm and 4.7 mm, which connects a first plateau at $n_{e1}=5\times10^{17}$ cm$^{-3}$ to a second plateau at $n_{e2}=2\times10^{17}$ cm$^{-3}$. Confining the injection to the ramp suppresses continuous self-injection along the channel and keeps the accelerated bunch free of dark current.

Figure \ref{FigLWFA} shows the wake structure and the bunch properties after 100 mm of laser propagation, where panel (a) gives the electron density of the wake together with the on-axis accelerating field $E_z$ at the same instant. A Gaussian fit to the spectral peak gives a central energy of 3.25 GeV with an r.m.s. energy spread of 1.85 \%, and the bunch carries 246 pC. Taking all macroparticles into account raises the mean energy to 3.49 GeV and the r.m.s. energy spread to 10.9 \% because of the low-energy tail visible in Fig. \ref{FigLWFA}b. The r.m.s. bunch duration is 9.2 fs with a FWHM of 4.6 fs, which gives a peak current of 25.8 kA. At the channel exit the bunch has an r.m.s. size of 0.76 $\mu$m and 0.53 $\mu$m, an r.m.s. divergence of 0.495 mrad and 0.345 mrad and a normalized emittance of 2.65 mm mrad and 1.23 mm mrad in $x$ and $y$.

The divergence at the channel exit lies at the sub-milliradian to few-milliradian scale typical of LWFA beams and still exceeds the angular acceptance for channeling at 3 GeV. We, therefore, transport the bunch through an active plasma lens, which provides an azimuthally symmetric focusing gradient at the kilotesla-per-metre level and captures the beam within a few centimetres of the plasma exit\cite{Tilborg2015,Lindstrom2018,Integrated2023,Gustafsson2024,Radiography2024,Drobniak2025}. Linear active plasma lenses built around a 500 $\mu$m diameter capillary have been operated at kiloampere discharge currents and reach focusing gradients of 3.6 kT m$^{-1}$\cite{Sjobak2021}, so the gradient required here stays within the demonstrated range. The lens parameters and the beamline layout are specified in \textbf{Methods} and place the crystal 250 mm from the source.

Figure \ref{FigAPL}a shows the evolution of the envelope along this beamline. At the crystal the bunch reaches an r.m.s. size of 2.02 $\mu$m and 1.41 $\mu$m, an r.m.s. divergence of 344 $\mu$rad and 240 $\mu$rad in $x$ and $y$, and bunch pulse duration of 8.6 fs (FWHM), as shown in Fig. \ref{FigAPL}b-d.

These values reproduce the 3 GeV entry of Table \ref{Table0}, which assumes a 2 $\mu$m r.m.s. size, a 0.4 mrad r.m.s. divergence and 10 fs bunch duration (FWHM). The resulting sub-milliradian divergence is comparable to the acceptance set by characteristic angles for the manifestation of coherent effects in oriented crystals for the energies considered in this paper, as will be described below.

The simulations show that the beam properties assumed for the 3 GeV case are attainable in a single LWFA stage, and we therefore adopt this specification throughout the paper. Because the 300 MeV and 1 GeV beams have already been extensively validated experimentally and through simulation in the literature cited above, we used beam parameters taken from those reports for these two energies.

\subsection*{Beam influence to coherent effects}

The electron-beam energy spread is particularly important when quasi-monochromatic radiation is required. In this work, however, we focus on maximizing the photon yield over a broad energy range to enhance the subsequent production of secondary particles. We therefore consider electrons incident along a major $\langle111\rangle$ axis of a W crystal, where strong coherent effects produce intense broadband radiation. The possibility of generating quasi-monochromatic radiation is discussed separately at the end of this section.

The interaction of electrons with an oriented crystal and the resulting radiation emission are characterized by two relevant angular scales: the critical channeling (Lindhard) angle, $\theta_L=\sqrt{2U_0/pv}$, and the characteristic angle splitting dipole-like and synchrotron-like radiation regimes, $\theta_v=U_0/m_ec^2$. Here, $U_0$ is the depth of the transverse potential well, $p$ and $v$ are the particle momentum and velocity, and $m_e$ is the electron mass. The Lindhard angle defines the approximate angular acceptance for channeling, $\theta<\theta_L$, and decreases with increasing particle energy. In contrast, the angle $\theta_v$ is energy independent and characterizes the radiation regime: particle trajectories at angles $\theta\gg\theta_v$ are associated with dipole-like emission, whereas $\theta\ll\theta_v$ corresponds to the synchrotron-like regime. The transition from $\theta_L>\theta_v$ to $\theta_L<\theta_v$ indicates a shift of channeling radiation from the dipole-like towards the synchrotron-like regime. Crystal alignment and beam angular divergence are consequently key parameters governing both channeling and radiation emission. For reference, the beam divergences listed in Table \ref{Table0} are compared with $\theta_L$ and $\theta_v$. For the conditions considered here, this transition occurs between 300 MeV and 1 GeV. Even at 300 MeV, however, $\theta\sim\theta_v$, indicating that the radiation is already outside the purely dipole-like regime. This is also supported by comparing $\theta_L$ with the characteristic radiation cone angle $\theta_c$ inversely proportional to the Lorenz-factor of electrons $\gamma$, $\theta_c=1/\gamma$. This angle remains smaller than $\theta_L$ for all three beam energies, whereas the opposite relation is expected in the dipole-like regime. We therefore expect predominantly broadband emission for all three beam energies considered.

\subsection*{Axial channeling along <111> in a W crystal}

\begin{figure*}[ht]
\centering
\includegraphics[width=0.49\textwidth]{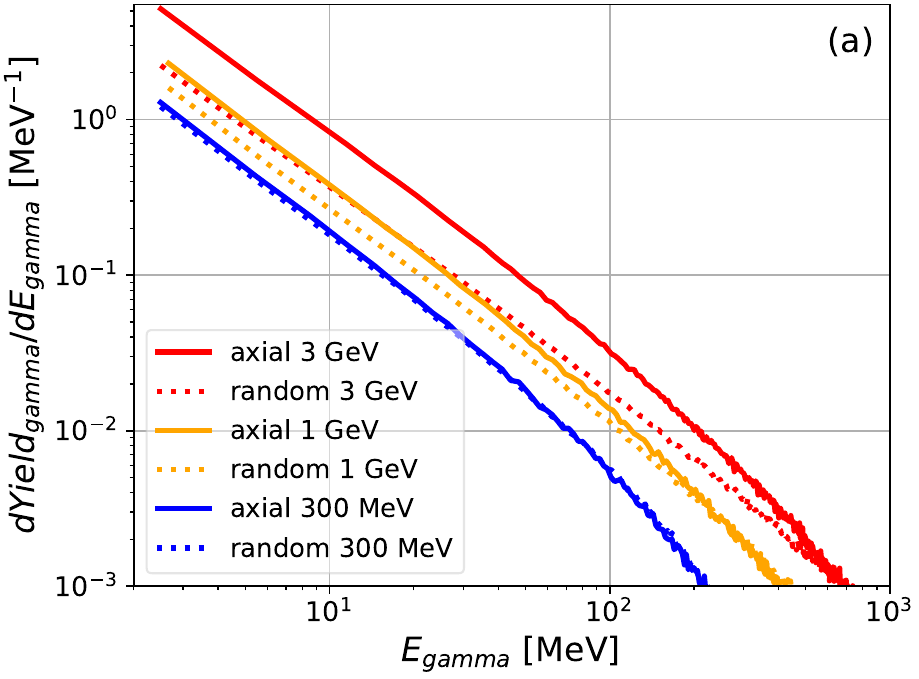}%
\includegraphics[width=0.49\textwidth]{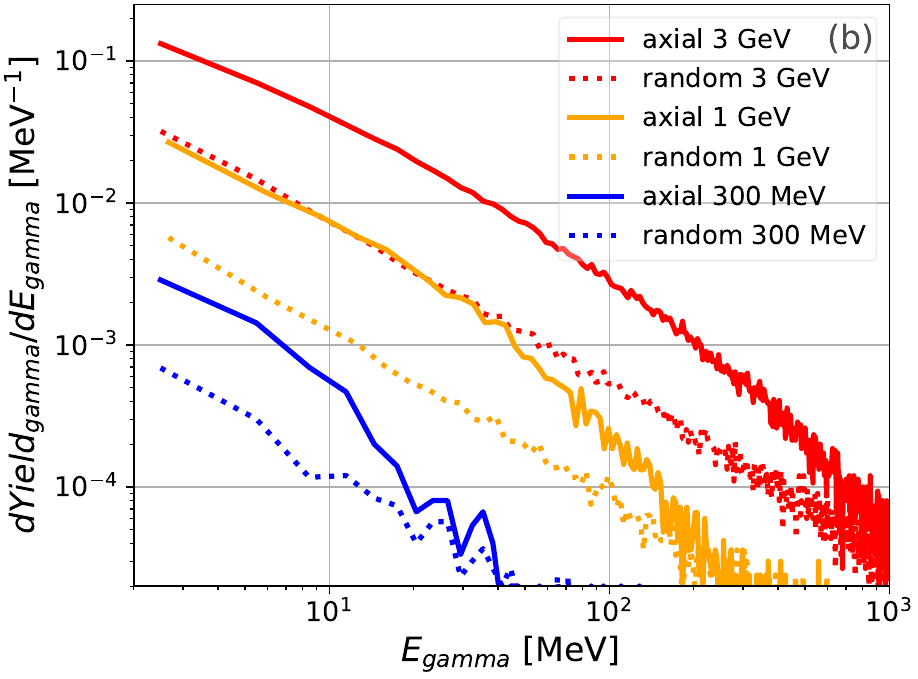}

\includegraphics[width=0.49\textwidth]{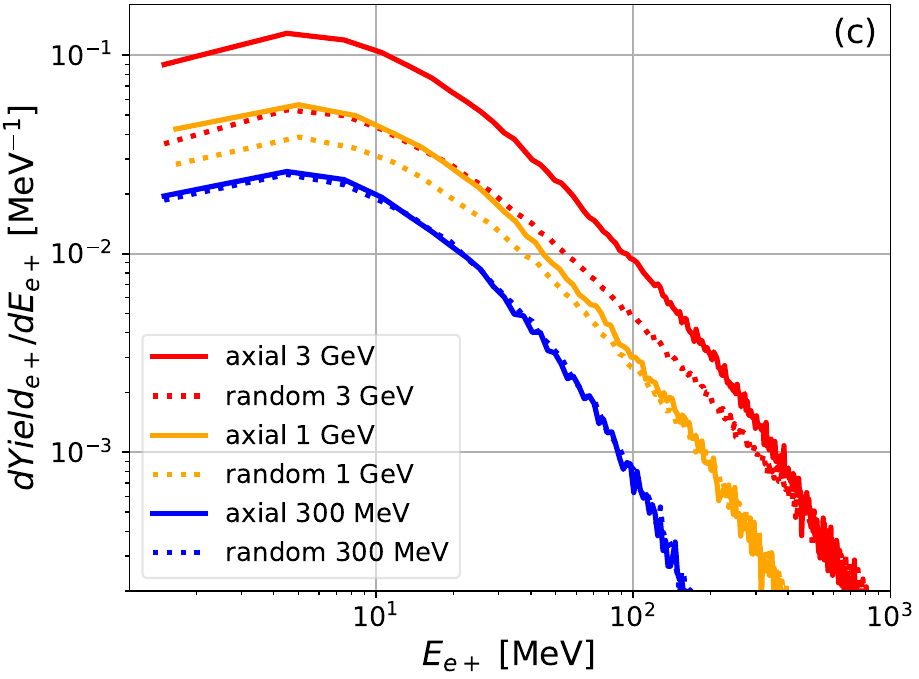}%
\includegraphics[width=0.49\textwidth]{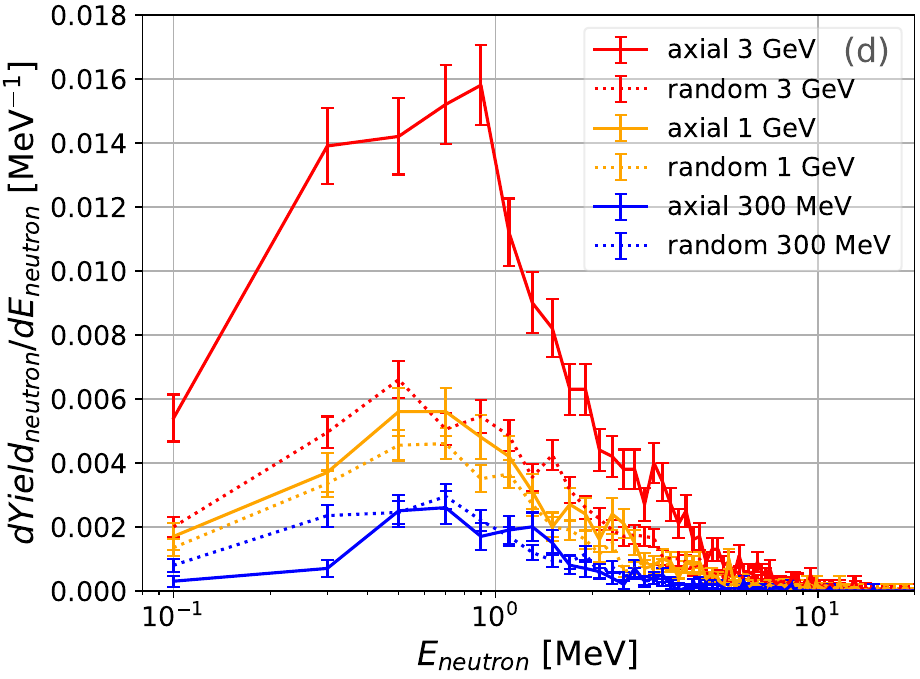}

\caption{Simulated spectra of $\gamma$-rays (a), collimated $\gamma$-rays within the radial angle of 1 mrad (b), positrons (c) and neutrons (d) for the simulation parameters from Table \ref{Table0} for a 5 mm thick W crystal either aligned along the $\langle111\rangle$ axis or randomly oriented. The spectra are normalized to the total particle production yield. $\gamma$-rays were selected with the energies > 1 MeV. Errorbars represent statistical uncertainties  (only on plots where they are important).}
\label{Fig2}
\end{figure*}

\begin{figure*}[ht]
\centering
\includegraphics[width=0.49\textwidth]{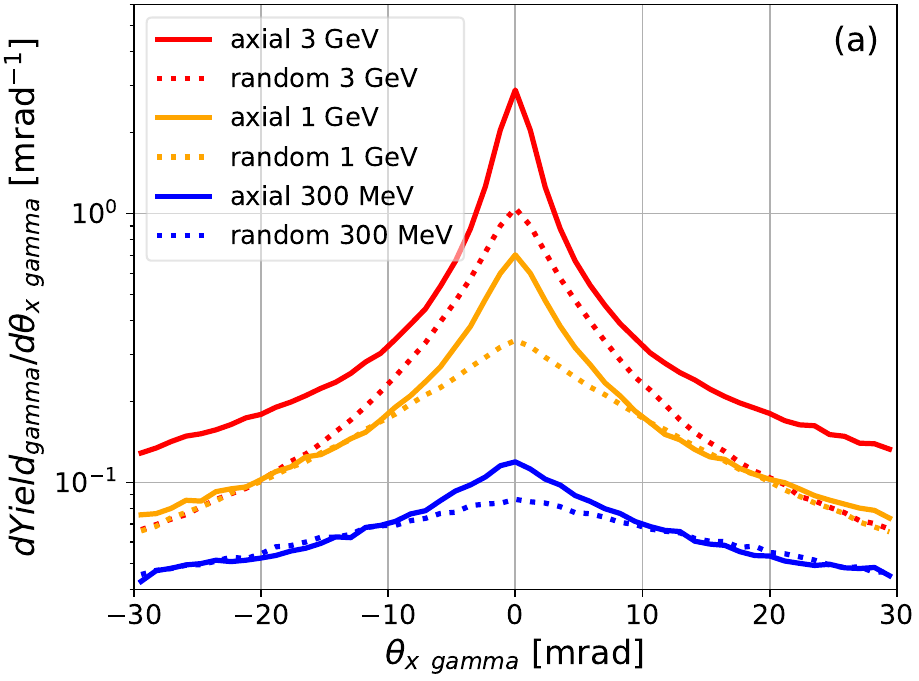}%
\includegraphics[width=0.49\textwidth]{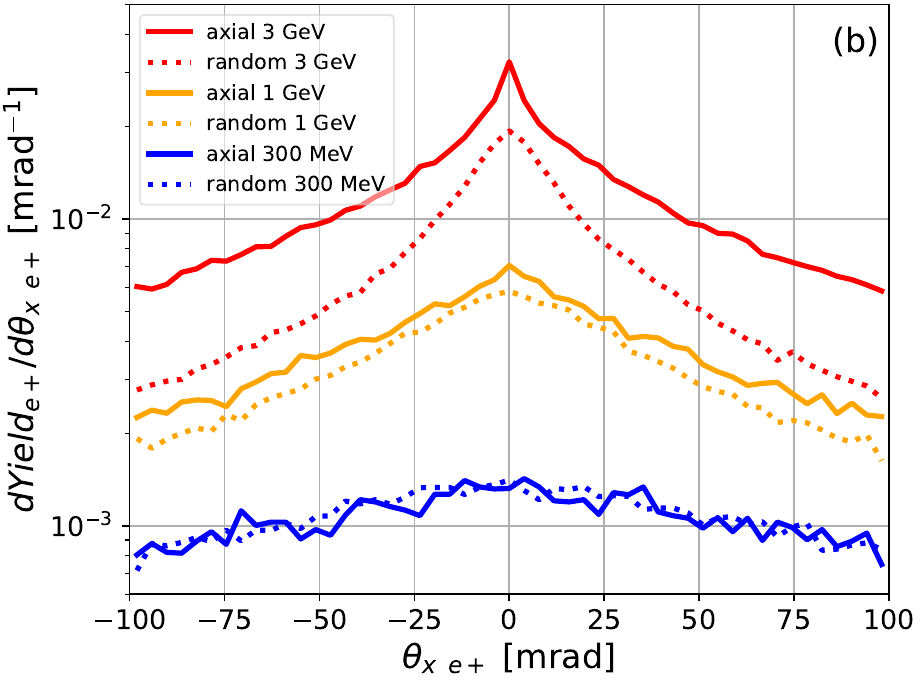}

\includegraphics[width=0.49\textwidth]{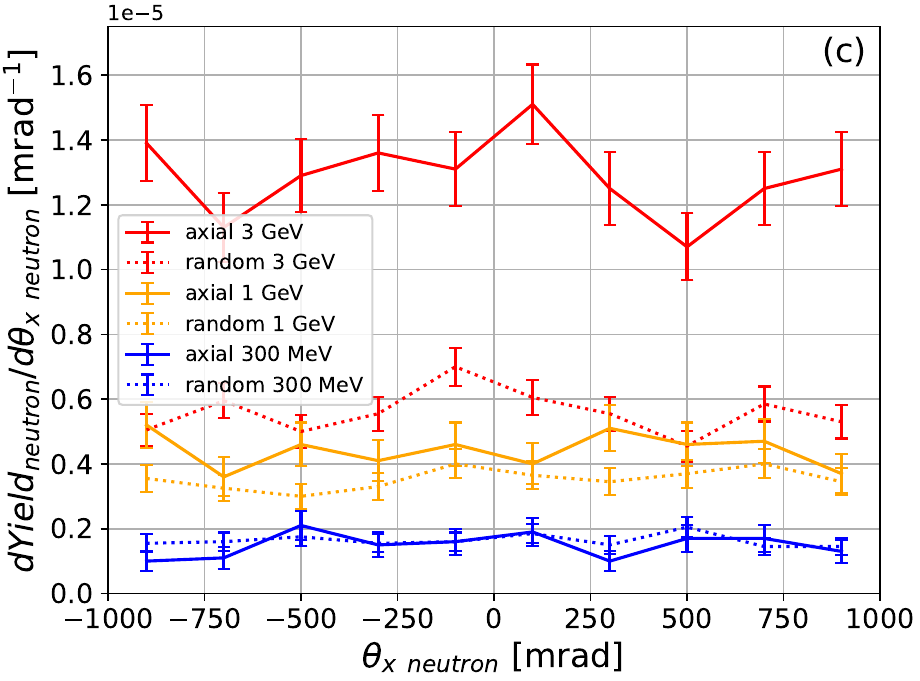}%
\includegraphics[width=0.49\textwidth]{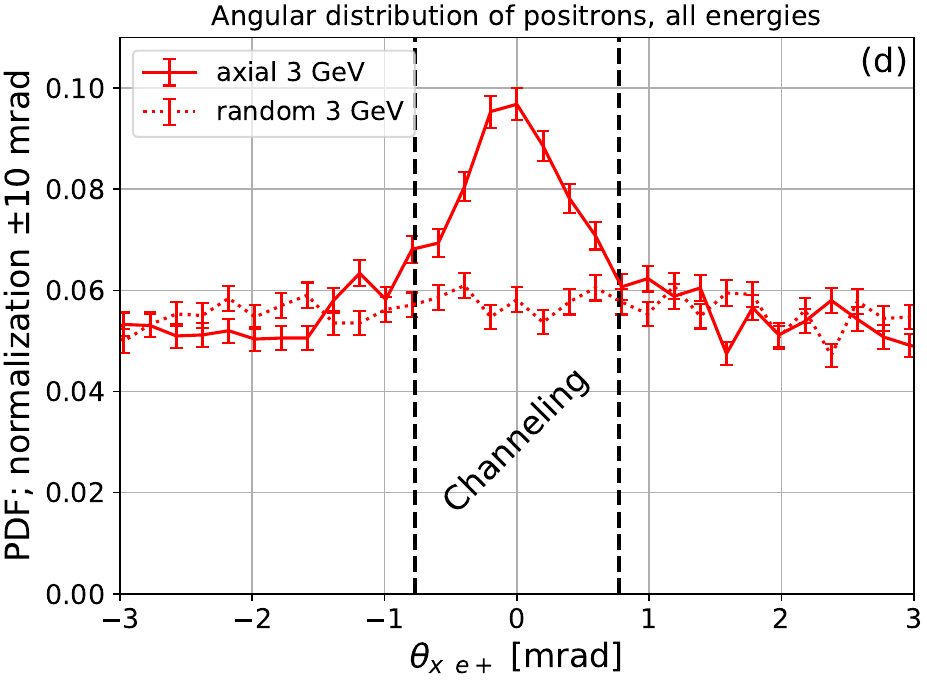}

\caption{Simulated angular distributions of $\gamma$-rays (a), positrons (b) and neutrons (c). (d) represent the PDF angular distribution for the 3 GeV case to demonstrate the capture of produced positrons under the channeling conditions. The distribution is normalized within $\pm$10 mrad) to consider only the central part of the distribution. Vertical lines represent $\theta_L$ for 3 GeV - minimal possible critical channeling angle. The simulation parameters are the same as those used in Fig. \ref{Fig2}. Errorbars represent statistical uncertainties (only on plots where they are important).}
\label{Fig21}
\end{figure*}

\begin{figure}[t]
\centering
\:\:\:\includegraphics[
  width=0.9\linewidth,
  trim=0 37 0 0,
  clip
]{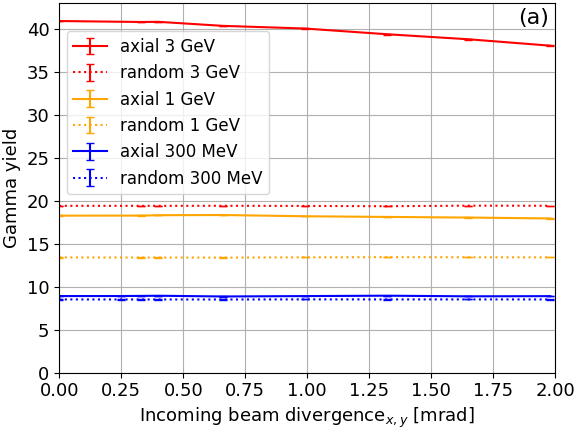}
\includegraphics[width=0.93\columnwidth]{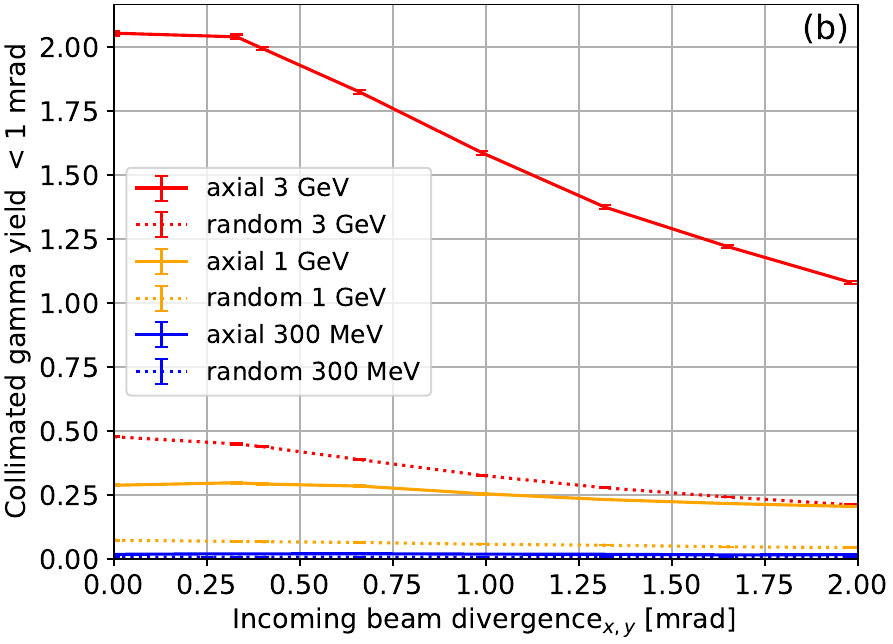}

\caption{$\gamma$-ray yields as a function of beam angular divergence for uncollimated (a) and collimated within the radial angle of 1 mrad (b) $\gamma$-rays. The remaining simulation parameters are identical to those used in Fig. \ref{Fig2}.}
\label{Fig3}
\end{figure}

\begin{figure}[ht]
\centering
\includegraphics[width=1\columnwidth]{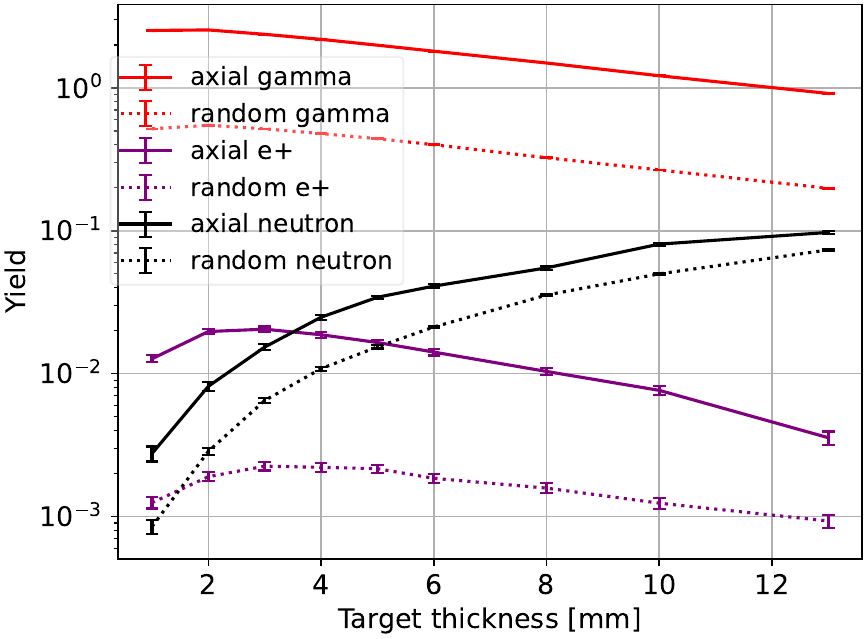}

\caption{
Particle yields as a function of target thickness for $\gamma$-rays and positrons within a radial angle of 1 mrad, and for uncollimated neutrons. The remaining simulation parameters are identical to those used in Fig. \ref{Fig2}.}
\label{Fig4}
\end{figure}

\begin{figure}[ht]
\centering
\includegraphics[width=1\columnwidth]{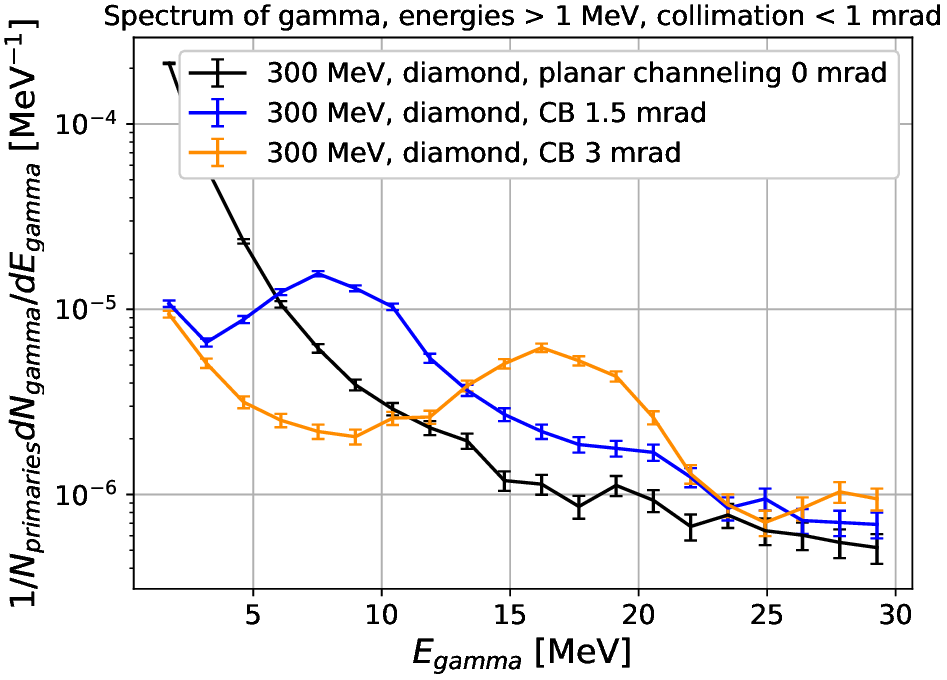}

\caption{Simulated spectra of coherent bremsstrahlung radiation produced by a 300 MeV beam (from Table \ref{Table0}) in a 10 $\mu$m-thick diamond crystal for planar channeling orientation along (110) planes and 2 orientations for CB. The collimation radial angle below 1 mrad was applied. The estimated brilliance for each of CB peaks for a 200 pC beam and a 10 fs pulse reaches $\sim3.5\cdot10^{20} \gamma/s/mm^2/mrad^2/0.1\% BW$.}
\label{Fig5}
\end{figure}

The simulations were carried out for both axial $\langle111\rangle$ alignment and a randomly oriented W crystal of 5 mm thickness and 20 mm width at three electron energies listed in Table \ref{Table0}. Fig. \ref{Fig2} shows the energy spectra of $\gamma$-rays, positrons and neutrons as well as the spectra of photons emitted within 1 mrad of the beam axis while angular distributions are presented in Fig. \ref{Fig21}.

\subsubsection*{$\gamma$-radiation}

The $\gamma$-ray spectra demonstrate a substantial enhancement of the photon yield over a broad energy range for the axially aligned crystal compared with the randomly oriented case at 1 and 3 GeV, while for 300 MeV it happens only in a central region angular distribution (compare Fig. \ref{Fig2}a-b and \ref{Fig21}a). 

The enhancement becomes more pronounced and happens for more energetic photons with increasing electron energy, as the radiation evolves towards the more intense synchrotron-like regime. At the same time, multiple scattering produces smaller angular deflections at higher energies, allowing electrons to remain aligned with the crystal axis over longer distances and thereby favoring coherent interactions. In our case, increasing the electron energy by a factor of $\sim3$ shifts the enhanced photon emission by approximately one order of magnitude in energy.

The enhancement is strongly angle dependent and is most pronounced close to the beam axis, as evident from both the collimated spectra - with angles below within a radial angle of 1 mrad (Fig. \ref{Fig2}b) and angular distributions (Fig. \ref{Fig21}a), both normalized to the total particle yield. Unlike 300 MeV, for 3 GeV, the enhancement remains significant over a wide range of radiation angles. The secondary electrons and positrons produced in the crystal contribute to this process.

Table \ref{Table1} summarizes the calculated photon yields for all cases considered. To estimate the total number of photons per laser pulse, these yields were scaled to an incoming beam charge of 200 pC, a typical value as discussed above. The results demonstrate the possibility to produce $\sim 10^{10}$ photons and up to $\sim 10^{9}$ collimated photons per pulse or up to $\sim 10^{22}-10^{24}$ photons per second for a 10 fs pulse (see Table \ref{Table0}). The latter demonstrates the advantage of using LWFA beams for an ultrafast radiation source. This is also confirmed by a brightness that can be estimated as a number of photons produced per unit of time per r.m.s. area and solid angle for radiation and per 0.1\% bandwidth - all the spectrum in our case ($dN_\gamma /dt dA d\Omega \cdot 10^{-3}$), reaching $\sim 10^{24} \gamma/s/mm^2/mrad^2$. 

Furthermore, the use of an oriented crystal increases the total photon yield by up to a factor of $\sim 2$ and the collimated photon yield by up to a factor of $\sim 5$, compared with a randomly oriented crystal under the conditions considered here. The peak brightness demonstrates even higher ratio above the factor of $\sim 6-8$ due to more intense radiation at lower angles caused by coherent effects in the crystal. Importantly, even at 300 MeV, such  substantial enhancement is preserved for collimated photons.

\begin{table*}[ht]
\centering
\begin{tabular}{|l|l|l|l|l|l|l|l|l|l|}
\hline
  &  \multicolumn{2}{c|}{$\gamma$} & \multicolumn{3}{c|}{$\gamma$ (< 1 mrad)} & \multicolumn{2}{c|}{e+} & \multicolumn{2}{c|}{neutrons}\\
\hline
case & \multicolumn{2}{c|}{yield} &\multicolumn{3}{c|}{yield} &\multicolumn{2}{c|}{yield} &\multicolumn{2}{c|}{yield} \\
\hline
axial 3 GeV& \multicolumn{2}{c|}{40.80$\pm$0.03} & \multicolumn{3}{c|}{1.993$\pm$0.007} & \multicolumn{2}{c|}{4.862$\pm$0.010}& \multicolumn{2}{c|}{(3.43$\pm$0.09)$10^{-2}$}\\
\hline
rnd 3 GeV & \multicolumn{2}{c|}{19.391$\pm$0.014} &  \multicolumn{3}{c|}{0.438$\pm$0.002}& \multicolumn{2}{c|}{2.401$\pm$0.005}& \multicolumn{2}{c|}{(1.49$\pm$0.04)$10^{-2}$}\\
\hline
axial 1 GeV & \multicolumn{2}{c|}{18.30$\pm$0.02} & \multicolumn{3}{c|}{0.292$\pm$0.003}& \multicolumn{2}{c|}{1.819$\pm$0.006} & \multicolumn{2}{c|}{(1.19$\pm$0.05)$10^{-2}$}\\
\hline
rnd 1 GeV & \multicolumn{2}{c|}{13.381$\pm$0.012} &\multicolumn{3}{c|}{(6.72$\pm$0.09)$10^{-2}$} & \multicolumn{2}{c|}{1.409$\pm$0.004}& \multicolumn{2}{c|}{(9.6$\pm$0.3)$10^{-3}$}\\
\hline
axial 300 MeV & \multicolumn{2}{c|}{8.932$\pm$0.014} & \multicolumn{3}{c|}{(1.97$\pm$0.07)$10^{-2}$}& \multicolumn{2}{c|}{0.656$\pm$0.004}& \multicolumn{2}{c|}{(4.1$\pm$0.3)$10^{-3}$}\\
\hline
rnd 300 MeV & \multicolumn{2}{c|}{8.499$\pm$0.010} & \multicolumn{3}{c|}{(6.4$\pm$0.3)$10^{-3}$}& \multicolumn{2}{c|}{0.655$\pm$0.003}& \multicolumn{2}{c|}{(4.6$\pm$0.3)$10^{-3}$}\\
\hline

case & 
$\displaystyle\gamma/pulse$ &
$\displaystyle\gamma/s$ &
$\displaystyle\gamma/pulse$ &
$\displaystyle\gamma/s$ &
$\displaystyle\gamma/s/mm^2/$ &
$\displaystyle e+/pulse$ &
$\displaystyle e+/s$ &
$\displaystyle n/pulse$ &
$\displaystyle n/s$ \\

 & $\times 10^{10}$ & $\times 10^{24}$ & $\times 10^{7}$ & $\times 10^{21}$ & $mrad^{2}/0.1\%BW$ & $\times 10^{9}$ & $\times 10^{23}$ & $\times 10^{7}$ & $\times 10^{21}$ \\
\hline
axial 3 GeV& $5.1$ & $5.1$ & $250$ & $250$ & $1.9 \cdot 10^{24}$ &$6.1$ & $6.1$ &$4.3$& $4.3$\\
\hline
rnd 3 GeV & $2.4$ & $2.4$  & $55$& $55$ & $3.1 \cdot 10^{23}$  & $3.0$& $3.0$ & $1.9$& $1.9$ \\
\hline
axial 1 GeV & $2.3$& $2.3$ & $36$ & $36$ & $9.9 \cdot 10^{22}$ & $2.3$ & $2.3$& $1.5$ & $1.5$\\
\hline
rnd 1 GeV & $1.7$ & $1.7$ & $8.4$ & $8.4$ & $1.2 \cdot 10^{22}$ & $1.8$ & $1.8$ & $1.2$ & $1.2$\\
\hline
axial 300 MeV & $1.1$ & $1.1$ & $2.5$ & $2.5$ & $1.2 \cdot 10^{21}$ & $0.8$ & $0.8$ & $0.5$ & $0.5$\\
\hline
rnd 300 MeV & $1.1$ & $1.1$ & $0.8$ & $0.8$ & $1.7 \cdot 10^{20}$ & $0.8$ & $0.8$ & $0.5$ & $0.5$\\
\hline
\end{tabular}
\caption{\label{Table1} Simulated total yields of $\gamma$-rays, positrons and neutrons over the full energy range, together with the yields of photons collimated within a radial angle of 1 mrad and the corresponding total numbers of particles produced per laser pulse ($particle/pulse$) and per unit of time during a 10 fs pulse ($particle/s$), scaled to an electron beam charge of 200 pC. For the collimated photon case the brightness was also calculated. The total yields correspond to the integrals of the spectra in Fig. \ref{Fig2} over the full energy range. $\gamma$-rays were selected with the energies > 1 MeV. The simulation parameters are the same as those used in Fig. \ref{Fig2}. Statistical uncertainties for yields are also reported.}
\end{table*}

\subsubsection*{positron and neutron production}

The energy spectra and angular distributions of the produced positrons and neutrons normalized to the total particle yield, are shown in Figs. \ref{Fig2}-\ref{Fig21}, similarly to $\gamma$-rays. Their production yields are summarized together with the photon yields in Table \ref{Table1}.

Positron production through gamma conversion shown in Fig. \ref{Fig2}c, \ref{Fig21}b follows a similar trend to the total photon yield, as expected from the pair-production process. The enhancement is considerably weaker at lower electron energies and becomes nearly negligible at 300 MeV. This is also evident from Table \ref{Table1}, which shows nearly identical positron yields for the axial and random crystal alignments, together with only a slight enhancement in the total $\gamma$-ray yield. At higher energies, however, the positron yield increases by up to a factor of $\sim 2$, consistent with the enhancement of the total photon yield.

Moreover, at 3 GeV, a pronounced enhancement of the positron yield is observed at small angles, which is not explained solely by the predominantly forward direction of the parent photons. Positrons produced within the angular acceptance for channeling can become confined in the transverse potential between atomic strings, reducing the growth of their angular divergence due to multiple scattering. This effect is illustrated in Fig. \ref{Fig21}d by the positron angular probability density, normalized within ±10 mrad to consider only the central part of the distribution. For the 3 GeV axial case, a clear increase in the positron population is observed near the center of the distribution, approximately within the characteristic $\theta_L$ scale. The value of $\theta_L$ given in Table \ref{Table0} corresponds to 3 GeV particles and therefore represents the smallest channeling acceptance relevant here; most produced positrons have lower energies and consequently larger Lindhard angles. However, positrons captured with sufficiently low transverse energy are more likely to remain channeled through the crystal and retain a narrow angular distribution, consistent with the enhanced population within the $\sim mrad$ angular range observed in Fig. \ref{Fig21}d.

Neutrons produced through photonuclear reactions also follow the trend in the photon yield, as shown in Fig. \ref{Fig2}d and Table \ref{Table1}. A significant increase of neutron production is observed for 3 GeV in a broad energy range. At the same time the angular distribution of neutrons is nearly uniform due to the kinematics of photonuclear reactions and subsequent interactions within the target (Fig. \ref{Fig21}d). As for positrons, the enhancement decreases with electron energy and becomes nearly negligible at 300 MeV.

As for the $\gamma$-rays, Table \ref{Table1} summarizes the estimated total number of particles produced per laser pulse, scaled to a beam charge of 200 pC. The results indicate yields of up to $\sim 10^{9}$ positrons and $\sim 10^{7}$ neutrons per pulse ($\sim 10^{23}$ and $\sim 10^{21}$ particles per second within a 10 fs pulse, respectively), with the axial alignment nearly doubling both yields at 3 GeV compared with the randomly oriented case.

\subsubsection*{Parameter scans}

To assess the effect of beam angular divergence on coherent interactions and particle yields, we performed simulations for divergences from 0 to 2 mrad (Fig. \ref{Fig3}), covering the idealized limit and the experimentally relevant $\sim$mrad range\cite{Tilborg2015,Integrated2023,Gustafsson2024}. The simulations in Fig. \ref{Fig3}a demonstrate only a weak dependence of the $\gamma$-ray yield over the angular range considered, since most of this range lies within or close to the characteristic channeling angle $\theta_L$, allowing coherent interactions to remain effective. A similar trend is expected for secondary-particle production, as the corresponding yields are approximately proportional to the amount of produced photons.

In contrast, for collimated $gamma$-rays, the dependence on the electron-beam angular divergence becomes stronger as shown in Fig. \ref{Fig3}b. This is mainly because electrons propagating at larger angles emit photons predominantly along their instantaneous direction of motion, so a larger fraction of the radiation falls outside the collimator acceptance. In the dipole-like radiation regime, this sensitivity would be more pronounced because the spectral features vary significantly with the observation angle. Therefore, although the few-mrad angular divergence typically produced by LWFA beams (see \textbf{Methods}) may still allow coherent radiation effects to be exploited, lower divergence is preferable for maximizing their efficiency, motivating the use of a plasma lens.

To investigate the effect of crystal thickness, we performed a thickness scan of particle yields for the highest-energy case of 3 GeV, as shown in Fig. \ref{Fig4}. For $\gamma$-rays and positrons, we consider the collimated case, as previously, within a radial angle of 1 mrad. For positrons, this angular range is comparable in order of magnitude to the acceptance of beam-capture systems, such as solenoid lenses or adiabatic matching devices\cite{Bandiera2022Pair,Alharthi2025}.

The photon and positron yields reach a maximum at a certain target thickness, reflecting the development of the electromagnetic shower through successive bremsstrahlung emission and electron–positron pair production processes. Unlike the case without angular selection, where the absolute maximum yields remain comparable between the axial and randomly oriented crystals, the collimated yields exhibit an enhancement by a factor of $\sim5$, consistent with the collimated photon enhancement shown in Table \ref{Table1}. This enhancement remains significant over the entire investigated thickness range and reflects the pronounced angular dependence of coherent radiation. 

The previously selected crystal thickness of 5 mm is not the most optimal for the 1 mrad collimation considered here. However, the optimal thickness shifts towards larger values for wider collimation angles, as particles with larger accumulated angular divergence contribute to the accepted beam, resulting in an optimum thickness exceeding 5 mm. Thus, the crystal thickness should be optimized for the specific beam parameters and collimation requirements.

Unlike $\gamma$-rays and positrons, neutrons are produced with a broad angular distribution; therefore, no angular collimation is applied and the total yield is considered. The neutron yield remains higher for the axially aligned crystal over almost the entire investigated thickness range. This results from the conversion of enhanced $\gamma$-ray emission into neutrons through photonuclear reactions: higher photon yields directly translate into increased neutron production.

\subsection*{Coherent bremsstrahlung in a diamond crystal}

Coherent bremsstrahlung (CB) occurs when electrons traverse an oriented crystal outside the channeling regime, $\theta>\theta_L$, while remaining in the dipole regime, $\theta\gg\theta_v$ (see Fig. \ref{Fig1}). The radiation spectrum can exhibit pronounced coherent peaks, which can be further selected by collimation to produce quasi-monochromatic photon beams. Low-Z crystals such as diamond and Si are commonly used for this purpose because reduced multiple scattering and a weaker incoherent bremsstrahlung background help to preserve the coherent spectral features. Their high crystal quality and well-established fabrication technologies, largely developed by the semiconductor industry, provide additional practical advantages over high-Z materials such as W.

The main advantage of CB is the possibility of tuning the energy of the radiation peaks through crystal alignment. Varying the incidence angle modifies the distance between successive crystal-plane crossings along the electron trajectory and hence the effective undulation period. To demonstrate this effect, we performed simulations with the Geant4 G4ChannelingFastSimModel using the 300 MeV beam parameters from Table \ref{Table0} and the Geant4 setup described above (see also \textbf{Methods}). A 10 $\mu$m-thick diamond crystal was aligned along (110) planes and then inclined with respect to the incident beam direction. The emitted photons were collimated within 1 mrad, below the characteristic radiation cone angle $\theta_c$, to select the narrow angular region where the coherent spectral features are most pronounced. The results shown in Fig. \ref{Fig5} demonstrate the shift of the radiation peak from channeling-dominated emission at 0 mrad, to CB at 1.5 and 3 mrad. At larger incidence angles, the peak not only shifts towards higher photon energies but also becomes broader, as beam divergence and multiple scattering increasingly contribute over the characteristic angular scale of the undulating motion. The total $\gamma$-ray yield in the peaks is $\sim10^{-4}$ which scales to $\sim10^{5}$ per pulse for a 200 pC electron beam. 

Considering a 10 fs pulse, and selecting the $\gamma$-rays within the collimated angle of 1 mrad and within the radiation peak, one can estimate the radiation source brilliance. For each of CB peaks from Fig. \ref{Fig5} the brilliance reaches $\sim3.5\cdot10^{20} \gamma/s/mm^2/mrad^2/0.1\% BW$.

The yield and spectral bandwidth can be further optimized by adjusting the crystal thickness and beam angular divergence, providing a trade-off between intensity and spectral selectivity. Furthermore, increasing the beam energy improves the photon source performance through reduced multiple scattering and enhanced coherent emission.


\section*{Discussion}

Our simulations demonstrate the potential of an LWFA-driven crystal-based source for generating $\gamma$-rays, positrons and neutrons. The proposed scheme enables both intense broadband $\gamma$-ray emission through axial channeling and quasi-monochromatic emission through coherent bremsstrahlung. Crystal orientation substantially enhances the photon and secondary-particle yields compared with a randomly oriented target. This enhancement is strongly angle dependent and is most pronounced within the forward, $\sim$ mrad-scale emission cone.

The $\gamma$-ray yield is significantly enhanced over the entire electron-energy range considered in this paper, from 300 MeV to 3 GeV, which is accessible with modern tens-of-TW to PW-class laser systems. The concept could also be extended towards lower energies, from a few to tens of MeV, where discrete quantum transitions of channeled particles can produce photons in the X-ray range\cite{Baier1998,Bandiera2021}, potentially allowing operation with smaller and more accessible laser systems. At the opposite extreme, tens-of-GeV electron beams, approaching the energies explored in oriented-crystal experiments at SLAC FACET and CERN\cite{Bandiera2023,Bandiera2025}, may become accessible with next-generation multi-PW and 10-PW-class laser systems.

Coherent bremsstrahlung can provide quasi-monochromatic and linearly polarized $\gamma$-ray beams. Combined with the $\mu$m-scale source size characteristic of LWFA-driven systems, this can provide high brilliance, exceeding $10^{20} \gamma/s/mm^2/mrad^2/0.1\% BW$ in our case, and favorable coherence properties. This brilliance as well as the energy range of radiation is already comparable with the results obtained with a LWFA-driven Thomson Scattering source\cite{Sarri2014} for similar beam energy range to our case (300-600 MeV). The brilliance can be further optimized by the adjustment of the crystal thickness, beam energy and other beam parameters as a subject of our future study. The spectral bandwidth can also be further reduced by angular collimation, at the expense of photon intensity.

Such beams are promising for photonuclear studies, nuclear resonance fluorescence, isotope-selective imaging and non-destructive inspection\cite{Lan2023,Feng2023,Giubega2025}. The tunability of the coherent peak through crystal alignment further allows the photon energy to be matched to specific nuclear transitions or application requirements without any other change of the experimental setup.

Axial channeling in high-Z crystals provides a complementary regime, optimized for high photon yield rather than spectral selectivity. Modern laser facilities operating with a repetition rate of $\sim 10^0-10^1$ Hz allow one to reach the photon flux of $\sim10^{11}-10^{12}$ $\gamma$/s. The intense broadband $\gamma$-ray emission is particularly attractive for applications requiring high photon flux over a wide energy range, including photonuclear reactions, strong-field QED studies, high-energy radiography, radiation-damage studies, laboratory astrophysics and secondary-particle production\cite{Cole2018,Mirzaie2024,Tsai2026,Coated2026,Glinec2005}.

In particular, the proposed source can provide a compact route towards intense positron and, at higher energies, muon sources. Positrons generated through photon-induced pair production are of interest for antimatter physics, positron annihilation spectroscopy and future lepton collider concepts. In particular, this approach could generate  $\sim10^{10}-10^{11}$ positrons/s - 2-3 orders of magnitude fewer than the conventional positron source proposed for FCC-ee \cite{FCCeeInjector}, while potentially being significantly less expensive because it does not require a multi-GeV electron linac. Furthermore, the peak current of positrons, following from the peak production rate $\sim10^{23}$ e+/s, exceeds  $\sim10$ kA within a $\sim10$ fs pulse, which is not possible with any conventional accelerator technique.

At higher electron energies, the same principle can be extended towards muon production through photonuclear and Bethe–Heitler processes in high-Z converters. Recent experiments have demonstrated directional muon production using multi-GeV LWFA electron beams interacting with high-Z targets, highlighting the potential of laser-driven compact sources for applications such as muography and high-energy physics\cite{Terzani2025,Calvin2026,Zhang2025Muon}.

The enhanced photon yield can also be exploited for compact neutron production through photonuclear reactions, producing up to $\sim10^{8}-10^{9}$ neutrons/s with a peak production rate up to $\sim10^{21}$ neutrons/s within a 10 fs pulse. The resulting neutrons have a broad angular distribution characteristic of photonuclear production, in contrast to the strongly forward-directed photon and positron emission. Potential uses include neutron imaging, materials characterization, activation analysis, isotope production and radiation-damage testing of electronics and aerospace components\cite{Feng2020,Vallieres2025,Kim2026,Scheuren2024}.

The use of an oriented crystal implies precise control of its alignment and position, which can be achieved using a high-precision goniometer. As discussed above, varying the alignment allows different regimes of coherent radiation to be selected, while the crystal position can be adjusted to control the particle fluence. Oriented crystals have demonstrated substantial radiation hardness in accelerator experiments. Degradation of channeling properties in Si crystals has typically been observed at accumulated fluences of order $10^{20}$-$10^{21}$ charged particles/cm$^2$, although the damage strongly depends on the particle type, energy and irradiation conditions [REFs]. For the few-$\mu$m beam sizes and $\sim$200 pC bunch charges considered here, the single-pulse electron fluence can reach $10^{15}$-$10^{16}$ electrons/cm$^2$. This is significantly lower the degradation limit, even though the tolerance of oriented crystals to such highly localized ultrashort irradiation requires dedicated thermal and radiation-damage studies. Any constraint can be mitigated by positioning the crystal further downstream to increase the beam spot size. Alternatively, when a small source size is desirable, for example to improve spatial coherence in imaging applications, the crystal can be translated transversely after a suitable number of shots to distribute the accumulated dose over a larger area.

\section*{Conclusion}

In summary, combining LWFA electron accelerators with the enhanced radiation efficiency of oriented crystals provides a route towards compact, intense sources of $\gamma$-radiation and positrons, neutrons and potentially, with potential extensions to X-rays and muons. Crystal orientation enables either intense broadband emission for high-flux radiation and secondary-particle production, or spectrally selective $\gamma$-ray emission through coherent bremsstrahlung. Further experimental studies will be important to validate the predicted yields, spectra and angular distributions, assess target robustness under intense LWFA beams, and develop source configurations optimized to specific applications.

\section*{Methods}

\subsection*{G4ChannelingFastSimModel}

Geant4\cite{Agostinelli2003} is a Monte Carlo toolkit for simulating the transport and interactions of particles in matter, widely used in accelerator and high-energy physics, medical physics and space science. It enables the modelling of complex experimental setups and incorporates a broad range of electromagnetic and nuclear processes, together with particle transport and detector response. The initial beam distribution can be specified and subsequently simulated on a particle-by-particle basis. Collective beam effects are therefore not considered, and the bunch charge does not enter directly into the simulations. Absolute particle rates for a given bunch charge are obtained by scaling the simulated yield per electron by the number of electrons in the bunch.

G4ChannelingFastSimModel\cite{Sytov2023,Geant4Manual} simulates coherent interactions of charged particles in oriented crystals, including channeling, channeling radiation and coherent bremsstrahlung. Particle trajectories are calculated in the averaged potential of crystallographic planes or atomic strings. The model also accounts for incoherent scattering and the effect of incoherent scattering suppression in oriented crystals\cite{Sytov2019}.

Photon emission is simulated with G4BaierKatkov, which is coupled to G4ChannelingFastSimModel and implements the Baier–Katkov quasiclassical formalism\cite{Baier1998,Sytov2019}. The formalism is derived within strong-field QED, accounting for non-linear strong-field effects. The radiation probability is calculated along the classical particle trajectories provided by G4ChannelingFastSimModel, while quantum recoil is included, allowing the emission of hard photons to be described. The method neglects the quantum nature of the transverse particle motion and is therefore applicable when the discrete channeling levels can be approximated by a quasi-continuous spectrum. For electrons, this condition is typically reached at energies of approximately 100 MeV and above\cite{Baier1998,Sytov2023}, placing all three electron energies considered in this paper within the applicability range of the model. Therefore, this value was adopted as the low-energy cut for electrons and positrons in the G4ChannelingFastSimModel. For tungsten, we additionally set the upper angular cut to $10\theta_L$ at the corresponding $e^\pm$ energy, which extends well beyond the channeling acceptance. For coherent bremsstrahlung in diamond, this cut was increased to $100\theta_L$ to include all relevant trajectories within the angular range of interest. Outside these energy and angular limits, the G4ChannelingFastSimModel is not activated, and particle transport is simulated using the standard Geant4 physics models. These cuts are necessary to keep the simulation time within reasonable limits.

The simulations were performed with Geant4 v11.4.0 using the G4ChannelingFastSimModel. The simulation application was based on the extended Geant4 example ch2 distributed with Geant4 v11.4.1. The G4ChannelingFastSimModel relevant to the present simulations is unchanged between these versions. We also used the G4CHANNELINGDATA 2.0 dataset included in Geant4, which provides a collection of crystal materials and crystallographic orientations, including diamond and W considered in this paper.

The main modification to the ch2 setup was the use of the FTFP\_BERT\_HP reference physics list to provide high-precision neutron transport and interactions. The selected physics also includes the electromagnetic and photonuclear processes relevant to $\gamma$-ray conversion and secondary particle production. The G4BaierKatkov parameters were kept at their default ch2 values. The remaining parameters are specified in the Results section to facilitate reproducibility of the simulations. The statistics of primary electrons used was 25k per configuration in Figs. \ref{Fig3}-\ref{Fig4} for the W crystal case, rising up to 50k for Figs. \ref{Fig2}-\ref{Fig21} and Table \ref{Table1}, and up to 100k for the random case per configuration in all plots. For diamond it reached 4M per curve in Fig. \ref{Fig5}.

The ch2 example also provides Python-based analysis tools, which were used as a starting point for the analysis of the simulated particle distributions and yields.

\subsection*{LWFA and beam transport simulations}

The electron beam used in the Results section was obtained in two steps. The acceleration stage was simulated with the quasi-cylindrical particle-in-cell (PIC) code FBPIC\cite{Lehe2016}, and the transport from the plasma exit to the crystal was calculated with the beam tracking code Wake-T\cite{FerranPousa2019}.

\textit{PIC simulation.} The simulations were performed in quasi-cylindrical geometry with $N_m=2$ azimuthal modes. The moving window was 100 $\mu$m long and extended to 100 $\mu$m in radius and was resolved by $2000\times800$ cells in the longitudinal and radial directions, which gives $\Delta z=50$ nm. The time step was $\Delta t=\Delta z/c$. Macroparticles were loaded with $2\times2\times4$ per cell in the longitudinal, radial and azimuthal directions. The driver was a linearly polarized Gaussian pulse with $a_0=2.0$, a spot size $w_0=45~\mu$m FWHM in intensity, a duration $\tau=40$ fs FWHM in intensity and $\lambda_0=800$ nm, focused at $z_f=4.5$ mm, which corresponds to a peak power of about 200 TW. The plasma had a parabolic transverse profile $n(r)=n_e+\Delta n\,(r/w_m)^2$ with the matched channel depth $\Delta n=1/(\pi r_e w_m^2)$, where $r_e$ is the classical electron radius. The matched spot size was $w_{m1}=50~\mu$m in the first density plateau and $w_{m2}=30~\mu$m in the second one. Along the propagation axis the on-axis density was constant at $n_{e1}=5\times10^{17}$ cm$^{-3}$ up to $z=4.5$ mm, decreased linearly to $n_{e2}=2\times10^{17}$ cm$^{-3}$ between $z=4.5$ mm and $z=4.7$ mm, and stayed constant up to the channel end at $z=100$ mm. The entrance ramp was 10 mm long. A second simulation with $n_{e1}=4\times10^{17}$ cm$^{-3}$ and the same $n_{e2}$ was performed for comparison.

\textit{Beam transport.} Transport from the plasma exit to the crystal was modelled with the particle tracking code Wake-T v0.9.1\cite{FerranPousa2019}. The source beam had a central energy of 3.25 GeV, an r.m.s. energy spread of 1.85 \%, a charge of 234 pC, r.m.s. sizes $\sigma_x=0.76~\mu$m and $\sigma_y=0.53~\mu$m, r.m.s. divergences $\sigma'_x=0.495$ mrad and $\sigma'_y=0.345$ mrad and an r.m.s. duration of 9.2 fs. It was represented by a single ensemble of $6\times10^4$ macroparticles sampled from the transverse covariance matrices of the PIC distribution and reused for every configuration. The line comprised a 50 mm drift, a 100 mm active plasma lens element and a 100 mm drift to the crystal. The lens field was that of a uniform axial current density, $B_\theta(r)=g\,r$ with $g=\mu_0 I/(2\pi R^2)$ and bore radius $R=250~\mu$m, so that the gradient $g=2468$ T m$^{-1}$ used here corresponds to $I=771$ A. Beam-driven wakefields were disabled and the macroparticles were advanced with a fourth-order Runge-Kutta integrator. Distributions were recorded at 166 stations along the 250 mm line. The bore aperture, the radial non-uniformity of the discharge current density and scattering in the helium fill lie outside this field model and were assessed with a semi-analytic thick lens and drift-kick model applied to the same ensemble, which reproduced the Wake-T envelope to within 0.14 \% r.m.s.

\bibliography{references}



\section*{Acknowledgements}

This work was supported by the INFN CSN5 CORAL and Geant4INFN projects. We acknowledge ISCRA for awarding this project access to the LEONARDO supercomputer, owned by the EuroHPC Joint Undertaking, hosted by CINECA (Italy). This research work was supported by the National Research Foundation of Korea (Grant Nos. RS-2025-24873264 and RS-2022-NR070321), and the InnoCORE program of the Ministry of Science and ICT (UNIST 1.260036.01).








\end{document}